\documentclass[conference,10pt,twocolumn,letter]{IEEEtran}
\IEEEoverridecommandlockouts
\usepackage{cite}
\usepackage[T1]{fontenc}
\usepackage{graphicx}
\usepackage{amssymb}
\usepackage{amsmath}
\usepackage{paralist}
\usepackage{subfigure}
\usepackage{booktabs} %\toprule \midrule \bottomrule
\usepackage{algorithm}
\usepackage{algorithmic}
\usepackage{amsthm}
\usepackage{multirow}
\usepackage{microtype} % 
\usepackage{tablefootnote}
\usepackage{balance}

\usepackage{amssymb}
\usepackage{amsfonts}
\usepackage{mathrsfs}
\usepackage{xspace}
\usepackage{bm}
\usepackage{upgreek}

\newcommand{\safemath}[2]{\newcommand{#1}{\ensuremath{#2}\xspace}}

\safemath{\bma}{\mathbf{a}}
\safemath{\bmb}{\mathbf{b}}
\safemath{\bmc}{\mathbf{c}}
\safemath{\bmd}{\mathbf{d}}
\safemath{\bme}{\mathbf{e}}
\safemath{\bmf}{\mathbf{f}}
\safemath{\bmg}{\mathbf{g}}
\safemath{\bmh}{\mathbf{h}}
\safemath{\bmi}{\mathbf{i}}
\safemath{\bmj}{\mathbf{j}}
\safemath{\bmk}{\mathbf{k}}
\safemath{\bml}{\mathbf{l}}
\safemath{\bmm}{\mathbf{m}}
\safemath{\bmn}{\mathbf{n}}
\safemath{\bmo}{\mathbf{o}}
\safemath{\bmp}{\mathbf{p}}
\safemath{\bmq}{\mathbf{q}}
\safemath{\bmr}{\mathbf{r}}
\safemath{\bms}{\mathbf{s}}
\safemath{\bmt}{\mathbf{t}}
\safemath{\bmu}{\mathbf{u}}
\safemath{\bmv}{\mathbf{v}}
\safemath{\bmw}{\mathbf{w}}
\safemath{\bmx}{\mathbf{x}}
\safemath{\bmy}{\mathbf{y}}
\safemath{\bmz}{\mathbf{z}}
\safemath{\bmzero}{\mathbf{0}}
\safemath{\bmone}{\mathbf{1}}

\bmdefine{\biad}{a}
\bmdefine{\bibd}{b}
\bmdefine{\bicd}{c}
\bmdefine{\bidd}{d}
\bmdefine{\bied}{e}
\bmdefine{\bifd}{f}
\bmdefine{\bigd}{g}
\bmdefine{\bihd}{h}
\bmdefine{\biid}{i}
\bmdefine{\bijd}{j}
\bmdefine{\bikd}{k}
\bmdefine{\bild}{l}
\bmdefine{\bimd}{m}
\bmdefine{\bind}{n}
\bmdefine{\biod}{o}
\bmdefine{\bipd}{p}
\bmdefine{\biqd}{q}
\bmdefine{\bird}{r}
\bmdefine{\bisd}{s}
\bmdefine{\bitd}{t}
\bmdefine{\biud}{u}
\bmdefine{\bivd}{v}
\bmdefine{\biwd}{w}
\bmdefine{\bixd}{x}
\bmdefine{\biyd}{y}
\bmdefine{\bizd}{z}

\bmdefine{\bixid}{\xi}
\bmdefine{\bilambdad}{\lambda}
\bmdefine{\bimud}{\mu}
\bmdefine{\bithetad}{\theta}
\bmdefine{\biphid}{\phi}
\bmdefine{\bideltad}{\delta}

\safemath{\bmia}{\biad}
\safemath{\bmib}{\bibd}
\safemath{\bmic}{\bicd}
\safemath{\bmid}{\bidd}
\safemath{\bmie}{\bied}
\safemath{\bmif}{\bifd}
\safemath{\bmig}{\bigd}
\safemath{\bmih}{\bihd}
\safemath{\bmii}{\biid}
\safemath{\bmij}{\bijd}
\safemath{\bmik}{\bikd}
\safemath{\bmil}{\bild}
\safemath{\bmim}{\bimd}
\safemath{\bmin}{\bind}
\safemath{\bmio}{\biod}
\safemath{\bmip}{\bipd}
\safemath{\bmiq}{\biqd}
\safemath{\bmir}{\bird}
\safemath{\bmis}{\bisd}
\safemath{\bmit}{\bitd}
\safemath{\bmiu}{\biud}
\safemath{\bmiv}{\bivd}
\safemath{\bmiw}{\biwd}
\safemath{\bmix}{\bixd}
\safemath{\bmiy}{\biyd}
\safemath{\bmiz}{\bizd}

\safemath{\bmxi}{\bixid}
\safemath{\bmlambda}{\bilambdad}
\safemath{\bmmu}{\bimud}
\safemath{\bmtheta}{\bithetad}
\safemath{\bmphi}{\biphid}
\safemath{\bmdelta}{\bideltad}

\safemath{\bA}{\mathbf{A}}
\safemath{\bB}{\mathbf{B}}
\safemath{\bC}{\mathbf{C}}
\safemath{\bD}{\mathbf{D}}
\safemath{\bE}{\mathbf{E}}
\safemath{\bF}{\mathbf{F}}
\safemath{\bG}{\mathbf{G}}
\safemath{\bH}{\mathbf{H}}
\safemath{\bI}{\mathbf{I}}
\safemath{\bJ}{\mathbf{J}}
\safemath{\bK}{\mathbf{K}}
\safemath{\bL}{\mathbf{L}}
\safemath{\bM}{\mathbf{M}}
\safemath{\bN}{\mathbf{N}}
\safemath{\bO}{\mathbf{O}}
\safemath{\bP}{\mathbf{P}}
\safemath{\bQ}{\mathbf{Q}}
\safemath{\bR}{\mathbf{R}}
\safemath{\bS}{\mathbf{S}}
\safemath{\bT}{\mathbf{T}}
\safemath{\bU}{\mathbf{U}}
\safemath{\bV}{\mathbf{V}}
\safemath{\bW}{\mathbf{W}}
\safemath{\bX}{\mathbf{X}}
\safemath{\bY}{\mathbf{Y}}
\safemath{\bZ}{\mathbf{Z}}

\safemath{\bZero}{\mathbf{0}}
\safemath{\bOne}{\mathbf{1}}
\safemath{\bDelta}{\mathbf{\Delta}}
\safemath{\bLambda}{\mathbf{\UpLambda}}
\safemath{\bPhi}{\mathbf{\Upphi}}
\safemath{\bSigma}{\mathbf{\Upsigma}}
\safemath{\bOmega}{\mathbf{\Upomega}}
\safemath{\bTheta}{\mathbf{\Uptheta}}

\bmdefine{\biAd}{A}
\bmdefine{\biBd}{B}
\bmdefine{\biCd}{C}
\bmdefine{\biDd}{D}
\bmdefine{\biEd}{E}
\bmdefine{\biFd}{F}
\bmdefine{\biGd}{G}
\bmdefine{\biHd}{H}
\bmdefine{\biId}{I}
\bmdefine{\biJd}{J}
\bmdefine{\biKd}{K}
\bmdefine{\biLd}{L}
\bmdefine{\biMd}{M}
\bmdefine{\biOd}{N}
\bmdefine{\biPd}{O}
\bmdefine{\biQd}{P}
\bmdefine{\biRd}{R}
\bmdefine{\biSd}{S}
\bmdefine{\biTd}{T}
\bmdefine{\biUd}{U}
\bmdefine{\biVd}{V}
\bmdefine{\biWd}{W}
\bmdefine{\biXd}{X}
\bmdefine{\biYd}{Y}
\bmdefine{\biZd}{Z}

\bmdefine{\biDelta}{\Delta}
\bmdefine{\biLambda}{\Lambda}
\bmdefine{\biPhi}{\Phi}
\bmdefine{\biSigma}{\Sigma}
\bmdefine{\biOmega}{\Omega}
\bmdefine{\biTheta}{\Theta}

\safemath{\bimA}{\biAd}
\safemath{\bimB}{\biBd}
\safemath{\bimC}{\biCd}
\safemath{\bimD}{\biDd}
\safemath{\bimE}{\biEd}
\safemath{\bimF}{\biFd}
\safemath{\bimG}{\biGd}
\safemath{\bimH}{\biHd}
\safemath{\bimI}{\biId}
\safemath{\bimJ}{\biJd}
\safemath{\bimK}{\biKd}
\safemath{\bimL}{\biLd}
\safemath{\bimM}{\biMd}
\safemath{\bimN}{\biNd}
\safemath{\bimO}{\biOd}
\safemath{\bimP}{\biPd}
\safemath{\bimQ}{\biQd}
\safemath{\bimR}{\biRd}
\safemath{\bimS}{\biSd}
\safemath{\bimT}{\biTd}
\safemath{\bimU}{\biUd}
\safemath{\bimV}{\biVd}
\safemath{\bimW}{\biWd}
\safemath{\bimX}{\biXd}
\safemath{\bimY}{\biYd}
\safemath{\bimZ}{\biZd}

\safemath{\bimDelta}{\biDelta}
\safemath{\bimLambda}{\biLambda}
\safemath{\bimPhi}{\biPhi}
\safemath{\bimSigma}{\biSigma}
\safemath{\bimOmega}{\biOmega}
\safemath{\bimTheta}{\biTheta}

\safemath{\setA}{\mathcal{A}}
\safemath{\setB}{\mathcal{B}}
\safemath{\setC}{\mathcal{C}}
\safemath{\setD}{\mathcal{D}}
\safemath{\setE}{\mathcal{E}}
\safemath{\setF}{\mathcal{F}}
\safemath{\setG}{\mathcal{G}}
\safemath{\setH}{\mathcal{H}}
\safemath{\setI}{\mathcal{I}}
\safemath{\setJ}{\mathcal{J}}
\safemath{\setK}{\mathcal{K}}
\safemath{\setL}{\mathcal{L}}
\safemath{\setM}{\mathcal{M}}
\safemath{\setN}{\mathcal{N}}
\safemath{\setO}{\mathcal{O}}
\safemath{\setP}{\mathcal{P}}
\safemath{\setQ}{\mathcal{Q}}
\safemath{\setR}{\mathcal{R}}
\safemath{\setS}{\mathcal{S}}
\safemath{\setT}{\mathcal{T}}
\safemath{\setU}{\mathcal{U}}
\safemath{\setV}{\mathcal{V}}
\safemath{\setW}{\mathcal{W}}
\safemath{\setX}{\mathcal{X}}
\safemath{\setY}{\mathcal{Y}}
\safemath{\setZ}{\mathcal{Z}}
\safemath{\emptySet}{\varnothing}

\safemath{\colA}{\mathscr{A}}
\safemath{\colB}{\mathscr{B}}
\safemath{\colC}{\mathscr{C}}
\safemath{\colD}{\mathscr{D}}
\safemath{\colE}{\mathscr{E}}
\safemath{\colF}{\mathscr{F}}
\safemath{\colG}{\mathscr{G}}
\safemath{\colH}{\mathscr{H}}
\safemath{\colI}{\mathscr{I}}
\safemath{\colJ}{\mathscr{J}}
\safemath{\colK}{\mathscr{K}}
\safemath{\colL}{\mathscr{L}}
\safemath{\colM}{\mathscr{M}}
\safemath{\colN}{\mathscr{N}}
\safemath{\colO}{\mathscr{O}}
\safemath{\colP}{\mathscr{P}}
\safemath{\colQ}{\mathscr{Q}}
\safemath{\colR}{\mathscr{R}}
\safemath{\colS}{\mathscr{S}}
\safemath{\colT}{\mathscr{T}}
\safemath{\colU}{\mathscr{U}}
\safemath{\colV}{\mathscr{V}}
\safemath{\colW}{\mathscr{W}}
\safemath{\colX}{\mathscr{X}}
\safemath{\colY}{\mathscr{Y}}
\safemath{\colZ}{\mathscr{Z}}

\safemath{\opA}{\mathbb{A}}
\safemath{\opB}{\mathbb{B}}
\safemath{\opC}{\mathbb{C}}
\safemath{\opD}{\mathbb{D}}
\safemath{\opE}{\mathbb{E}}
\safemath{\opF}{\mathbb{F}}
\safemath{\opG}{\mathbb{G}}
\safemath{\opH}{\mathbb{H}}
\safemath{\opI}{\mathbb{I}}
\safemath{\opJ}{\mathbb{J}}
\safemath{\opK}{\mathbb{K}}
\safemath{\opL}{\mathbb{L}}
\safemath{\opM}{\mathbb{M}}
\safemath{\opN}{\mathbb{N}}
\safemath{\opO}{\mathbb{O}}
\safemath{\opP}{\mathbb{P}}
\safemath{\opQ}{\mathbb{Q}}
\safemath{\opR}{\mathbb{R}}
\safemath{\opS}{\mathbb{S}}
\safemath{\opT}{\mathbb{T}}
\safemath{\opU}{\mathbb{U}}
\safemath{\opV}{\mathbb{V}}
\safemath{\opW}{\mathbb{W}}
\safemath{\opX}{\mathbb{X}}
\safemath{\opY}{\mathbb{Y}}
\safemath{\opZ}{\mathbb{Z}}
\safemath{\opZero}{\mathbb{O}}
\safemath{\identityop}{\opI}

\safemath{\veca}{\bma}
\safemath{\vecb}{\bmb}
\safemath{\vecc}{\bmc}
\safemath{\vecd}{\bmd}
\safemath{\vece}{\bme}
\safemath{\vecf}{\bmf}
\safemath{\vecg}{\bmg}
\safemath{\vech}{\bmh}
\safemath{\veci}{\bmi}
\safemath{\vecj}{\bmj}
\safemath{\veck}{\bmk}
\safemath{\vecl}{\bml}
\safemath{\vecm}{\bmm}
\safemath{\vecn}{\bmn}
\safemath{\veco}{\bmo}
\safemath{\vecp}{\bmp}
\safemath{\vecq}{\bmq}
\safemath{\vecr}{\bmr}
\safemath{\vecs}{\bms}
\safemath{\vect}{\bmt}
\safemath{\vecu}{\bmu}
\safemath{\vecv}{\bmv}
\safemath{\vecw}{\bmw}
\safemath{\vecx}{\bmx}
\safemath{\vecy}{\bmy}
\safemath{\vecz}{\bmz}

\safemath{\veczero}{\bmzero}
\safemath{\vecone}{\bmone}
\safemath{\vecxi}{\bmxi}
\safemath{\veclambda}{\bmlambda}
\safemath{\vecmu}{\bmmu}
\safemath{\vectheta}{\bmtheta}
\safemath{\vecphi}{\bmphi}
\safemath{\vecdelta}{\bmdelta}

\safemath{\matA}{\bA}
\safemath{\matB}{\bB}
\safemath{\matC}{\bC}
\safemath{\matD}{\bD}
\safemath{\matE}{\bE}
\safemath{\matF}{\bF}
\safemath{\matG}{\bG}
\safemath{\matH}{\bH}
\safemath{\matI}{\bI}
\safemath{\matJ}{\bJ}
\safemath{\matK}{\bK}
\safemath{\matL}{\bL}
\safemath{\matM}{\bM}
\safemath{\matN}{\bN}
\safemath{\matO}{\bO}
\safemath{\matP}{\bP}
\safemath{\matQ}{\bQ}
\safemath{\matR}{\bR}
\safemath{\matS}{\bS}
\safemath{\matT}{\bT}
\safemath{\matU}{\bU}
\safemath{\matV}{\bV}
\safemath{\matW}{\bW}
\safemath{\matX}{\bX}
\safemath{\matY}{\bY}
\safemath{\matZ}{\bZ}
\safemath{\matzero}{\bmzero}

\safemath{\matDelta}{\bDelta}
\safemath{\matLambda}{\bLambda}
\safemath{\matPhi}{\bPhi}
\safemath{\matSigma}{\bSigma}
\safemath{\matOmega}{\bOmega}
\safemath{\matTheta}{\bTheta}

\safemath{\matidentity}{\matI}
\safemath{\matone}{\matO}

\safemath{\rnda}{A}
\safemath{\rndb}{B}
\safemath{\rndc}{C}
\safemath{\rndd}{D}
\safemath{\rnde}{E}
\safemath{\rndf}{F}
\safemath{\rndg}{G}
\safemath{\rndh}{H}
\safemath{\rndi}{I}
\safemath{\rndj}{J}
\safemath{\rndk}{K}
\safemath{\rndl}{L}
\safemath{\rndm}{M}
\safemath{\rndn}{N}
\safemath{\rndo}{O}
\safemath{\rndp}{P}
\safemath{\rndq}{Q}
\safemath{\rndr}{R}
\safemath{\rnds}{S}
\safemath{\rndt}{T}
\safemath{\rndu}{U}
\safemath{\rndv}{V}
\safemath{\rndw}{W}
\safemath{\rndx}{X}
\safemath{\rndy}{Y}
\safemath{\rndz}{Z}

\safemath{\rveca}{\bimA}
\safemath{\rvecb}{\bimB}
\safemath{\rvecc}{\bimC}
\safemath{\rvecd}{\bimD}
\safemath{\rvece}{\bimE}
\safemath{\rvecf}{\bimF}
\safemath{\rvecg}{\bimG}
\safemath{\rvech}{\bimH}
\safemath{\rveci}{\bimI}
\safemath{\rvecj}{\bimJ}
\safemath{\rveck}{\bimK}
\safemath{\rvecl}{\bimL}
\safemath{\rvecm}{\bimM}
\safemath{\rvecn}{\bimN}
\safemath{\rveco}{\bomO}
\safemath{\rvecp}{\bimP}
\safemath{\rvecq}{\bimQ}
\safemath{\rvecr}{\bimR}
\safemath{\rvecs}{\bimS}
\safemath{\rvect}{\bimT}
\safemath{\rvecu}{\bimU}
\safemath{\rvecv}{\bimV}
\safemath{\rvecw}{\bimW}
\safemath{\rvecx}{\bimX}
\safemath{\rvecy}{\bimY}
\safemath{\rvecz}{\bimZ}

\safemath{\rvecxi}{\bmxi}
\safemath{\rveclambda}{\bmlambda}
\safemath{\rvecmu}{\bmmu}
\safemath{\rvectheta}{\bmtheta}
\safemath{\rvecphi}{\bmphi}

\safemath{\rmatA}{\bimA}
\safemath{\rmatB}{\bimB}
\safemath{\rmatC}{\bimC}
\safemath{\rmatD}{\bimD}
\safemath{\rmatE}{\bimE}
\safemath{\rmatF}{\bimF}
\safemath{\rmatG}{\bimG}
\safemath{\rmatH}{\bimH}
\safemath{\rmatI}{\bimI}
\safemath{\rmatJ}{\bimJ}
\safemath{\rmatK}{\bimK}
\safemath{\rmatL}{\bimL}
\safemath{\rmatM}{\bimM}
\safemath{\rmatN}{\bimN}
\safemath{\rmatO}{\bimO}
\safemath{\rmatP}{\bimP}
\safemath{\rmatQ}{\bimQ}
\safemath{\rmatR}{\bimR}
\safemath{\rmatS}{\bimS}
\safemath{\rmatT}{\bimT}
\safemath{\rmatU}{\bimU}
\safemath{\rmatV}{\bimV}
\safemath{\rmatW}{\bimW}
\safemath{\rmatX}{\bimX}
\safemath{\rmatY}{\bimY}
\safemath{\rmatZ}{\bimZ}

\safemath{\rmatDelta}{\bimDelta}
\safemath{\rmatLambda}{\bimLambda}
\safemath{\rmatPhi}{\bimPhi}
\safemath{\rmatSigma}{\bimSigma}
\safemath{\rmatOmega}{\bimOmega}
\safemath{\rmatTheta}{\bimTheta}

\usepackage{amssymb}
\usepackage{amsfonts}
\usepackage{mathrsfs}
\usepackage{xspace}
\usepackage{bm}
\usepackage{fancyref}
\usepackage{textcomp}

\usepackage{multirow}
\usepackage{stmaryrd}

\newenvironment{textbmatrix}{	\setlength{\arraycolsep}{2.5pt}%
								\big[\begin{matrix}}{\end{matrix}\big]%
								\raisebox{0.08ex}{\vphantom{M}}}

\def\be{\begin{equation}}
\def\ee{\end{equation}}
\def\een{\nonumber \end{equation}}
\def\mat{\begin{bmatrix}}
\def\emat{\end{bmatrix}}
\def\btm{\begin{textbmatrix}}
\def\etm{\end{textbmatrix}}

\def\ba#1\ea{\begin{align}#1\end{align}}
\def\bas#1\eas{\begin{align*}#1\end{align*}}
\def\bs#1\es{\begin{split}#1\end{split}}
\def\bg#1\eg{\begin{gather}#1\end{gather}}
\def\bml#1\eml{\begin{multline}#1\end{multline}}
\def\bi#1\ei{\begin{itemize}#1\end{itemize}}

\newcommand{\lefto}{\mathopen{}\left}

\DeclareMathOperator*{\argmin}{arg\;min}		% arg min
\DeclareMathOperator{\Exop}{\opE}			% expectation operator
\newcommand{\Ex}[2]{\ensuremath{\Exop_{#1}\lefto[#2\right]}} 	% expectation
\safemath{\dirac}{\delta}					% Dirac delta
\safemath{\krond}{\dirac}					% Kronecker delta
\safemath{\upto}{\uparrow}
\safemath{\downto}{\downarrow}
\safemath{\iu}{j}							% imaginary unit
\safemath{\ev}{\lambda}						% eigenvalue
\safemath{\hilseqspace}{l^{2}}				% Hilbert sequence space
\newcommand{\banachfunspace}[1]{\setL^{#1}}	% Banach function space
\safemath{\hilfunspace}{\banachfunspace{2}}	% Hilbert function space
\safemath{\SNR}{\textit{SNR}} 				% signal to noise ratio
\safemath{\PAR}{\textit{PAR}} 				% signal to noise ratio
\safemath{\No}{N_0}							% noise spectral density
\safemath{\Es}{E_s}							% energy per symbol
\safemath{\Eb}{E_b}							% energy per bit
\safemath{\EbNo}{\frac{\Eb}{\No}}
\safemath{\EsNo}{\frac{\Es}{\No}}

\DeclareMathOperator{\CHop}{\ensuremath{\opH}} % channel operator
\safemath{\tvir}{\rndh_{\CHop}}				% time-varying impulse response
\safemath{\tvtf}{\rndl_{\CHop}}				% 	-''- transfer function
\safemath{\spf}{\rnds_{\CHop}}				% spreading function
\safemath{\bff}{H_{\CHop}}					% bi-freuqency function

\safemath{\ircf}{r_{h}}						% impulse response correlation fn.
\safemath{\tftvcf}{r_{s}}					% scattering function
\safemath{\tfcf}{r_{l}}						% time-frequency correlation fn.
\safemath{\bfcf}{r_{H}}						% bi-frequency correlation fn.

\safemath{\tcorr}{c_h}						% time-correlation function
\safemath{\scf}{c_{s}}						% spreading function
\safemath{\tfcorr}{c_{l}}					% transfer-function correlation
\safemath{\fcorr}{c_{H}}						% frequency-correlation function

\safemath{\mi}{I}							% mutual information
\safemath{\capacity}{C}						% capacity

\safemath{\normal}{\mathcal{N}}			% normal distribution
\safemath{\jpg}{\mathcal{CN}}			% jointly proper Gaussian
\safemath{\mchain}{\leftrightarrow}		% Markov chain
\safemath{\dB}{\,\mathrm{dB}}
\safemath{\dBm}{\,\mathrm{dBm}}
\safemath{\Hz}{\,\mathrm{Hz}}
\safemath{\kHz}{\,\mathrm{kHz}}
\safemath{\MHz}{\,\mathrm{MHz}}
\safemath{\GHz}{\,\mathrm{GHz}}
\safemath{\s}{\,\mathrm{s}}
\safemath{\ms}{\,\mathrm{ms}}
\safemath{\mus}{\,\mathrm{\text{\textmu}s}}
\safemath{\ns}{\,\mathrm{ns}}
\safemath{\ps}{\,\mathrm{ps}}
\safemath{\meter}{\,\mathrm{m}}
\safemath{\mm}{\,\mathrm{mm}}
\safemath{\cm}{\,\mathrm{cm}}
\safemath{\m}{\,\mathrm{m}}
\safemath{\W}{\,\mathrm{W}}
\safemath{\mW}{\, \mathrm{mW}}
\safemath{\J}{\,\mathrm{J}}
\safemath{\K}{\,\mathrm{K}}
\safemath{\bit}{\,\mathrm{bit}}
\safemath{\nat}{\,\mathrm{nat}}

\safemath{\define}{\triangleq}			% definition

\safemath{\equivalent}{\sim}
\safemath{\distas}{\sim}					% distributed according to
\safemath{\sdiff}{\Delta}				% symmetric set difference

\safemath{\reals}{\mathbb{R}}
\safemath{\positivereals}{\reals_{+}}
\safemath{\integers}{\mathbb{Z}}
\safemath{\posint}{\integers_{+}}
\safemath{\naturals}{\mathbb{N}}
\safemath{\posnaturals}{\naturals_{+}}
\safemath{\complexset}{\mathbb{C}}
\safemath{\rationals}{\mathbb{Q}}

\newcommand*{\fancyrefapplabelprefix}{app}		% Appendix
\newcommand*{\fancyrefthmlabelprefix}{thm}		% Theorem
\newcommand*{\fancyreflemlabelprefix}{lem}		% Lemma
\newcommand*{\fancyrefcorlabelprefix}{cor}		% Corollary
\newcommand*{\fancyrefdeflabelprefix}{def}		% Definition
\newcommand*{\fancyrefproplabelprefix}{prop}		% Proposition
\newcommand*{\fancyrefexmpllabelprefix}{exmpl}
\newcommand*{\fancyrefalglabelprefix}{alg}		% Algorithm
\newcommand*{\fancyreftbllabelprefix}{tbl}		% Algorithm

\frefformat{vario}{\fancyrefseclabelprefix}{Section~#1}
\frefformat{vario}{\fancyrefthmlabelprefix}{Theorem~#1}
\frefformat{vario}{\fancyreftbllabelprefix}{Table~#1}
\frefformat{vario}{\fancyreflemlabelprefix}{Lemma~#1}
\frefformat{vario}{\fancyrefcorlabelprefix}{Corollary~#1}
\frefformat{vario}{\fancyrefdeflabelprefix}{Definition~#1}
\frefformat{vario}{\fancyreffiglabelprefix}{Fig.~#1}
\frefformat{vario}{\fancyrefapplabelprefix}{Appendix~#1}
\frefformat{vario}{\fancyrefeqlabelprefix}{(#1)}
\frefformat{vario}{\fancyrefproplabelprefix}{Proposition~#1}
\frefformat{vario}{\fancyrefexmpllabelprefix}{Example~#1}
\frefformat{vario}{\fancyrefalglabelprefix}{Algorithm~#1}

 \newtheorem{thm}{Theorem}
\safemath{\dictab}{[\,\dicta\,\,\dictb\,]}

\safemath{\ysig}{\bmy}
\safemath{\ysighat}{\hat{\ysig}}
\safemath{\ysigdim}{M}
\safemath{\xsig}{\bmx}
\safemath{\xsigdim}{N}
\safemath{\nx}{n_x}
\safemath{\zsig}{\bmz}
\safemath{\zsigdim}{\ysigdim}
\safemath{\rsig}{\bmr}
\safemath{\Adict}{\bA}
\safemath{\Adicttilde}{\widetilde{\Adict}}
\safemath{\Adictdim}{\outputdim\times\xsigdim}
\safemath{\avec}{\bma}
\safemath{\avectilde}{\tilde{\avec}}
\safemath{\Bdict}{\bB}
\safemath{\Bdicttilde}{\widetilde{\Bdict}}
\safemath{\Cdict}{\bC}
\safemath{\cvec}{\bmc}
\safemath{\Ddict}{\bD}
\safemath{\Ddictdim}{\ysigdim\times\xsigdim}
\safemath{\dvec}{\bmd}
\safemath{\Ddicttilde}{\widetilde{\bD}}
\safemath{\Bonb}{\bB}
\safemath{\bvec}{\bmb}
\safemath{\Bonbdim}{\ysigdim\times\ysigdim}
\safemath{\noise}{\bmn}
\safemath{\noisedim}{\ysigim}
\safemath{\err}{\bme}
\safemath{\errdim}{\ysigdim}
\safemath{\errset}{\setE}
\safemath{\nerr}{n_e}
\safemath{\delop}{\bP_\errset}
\safemath{\delopc}{\bP_{{\errset}^c}}

\safemath{\cplxi}{\imath}
\safemath{\cplxj}{\jmath}
\safemath{\dict}{\matD}
\safemath{\inputdim}{N}		% number of columns of dictionary D
\safemath{\outputdim}{M}		%number of rows of dictionary D
\safemath{\sparsity}{S}	%sparsity
\safemath{\inputdimA}{{N_a}}	%total number of elements in dictionary A
\safemath{\inputdimB}{{N_b}}	%total number of elements in dictionary B
\safemath{\elemA}{{n_a}}	%number of elements chosen from dictionary A
\safemath{\elemB}{{n_b}}	%number of elements chosen from dictionary B
\safemath{\resA}{\matR_a}	%restriction map to elements of dictionary A
\safemath{\resB}{\matR_b}	%restriction map to elements of dictionary B
\safemath{\subD}{\matS} %subdictionary
\safemath{\subA}{\matS_a} %subdictionary part of A
\safemath{\subB}{\matS_b} %subdictionary part of B
\safemath{\dicta}{\matA} 	% first subdictionary
\safemath{\dictb}{\matB} 	% second subdictionary
\safemath{\hollowS}{H}
\safemath{\hollowA}{H_a}
\safemath{\hollowB}{H_b}
\safemath{\cross}{Z}
\safemath{\coh}{\mu_d}			% coherence dictionary
\safemath{\coha}{\mu_a}			% coherence first subdictionary
\safemath{\cohb}{\mu_b}			% coherence second subdictionary
\safemath{\mubs}{\nu}	%block sub-coherence
\safemath{\cohm}{\mu_m} %mutual coherence
\safemath{\dictset}{\setD}	% set of dictionaries
\safemath{\dictsetp}{\dictset(\coh,\coha,\cohb)}	% set of dictionaries parametrized
\safemath{\dictsetgen}{\dictset_\text{gen}}
\safemath{\dictsetgenp}{\dictsetgen(\coh)}
\safemath{\dictsetonb}{\dictset_\text{onb}}
\safemath{\dictsetonbp}{\dictsetonb(\coh)}

\safemath{\leftside}{U}
\safemath{\rightsideA}{R_a}
\safemath{\rightsideB}{R_b}

\safemath{\indexS}{\setI_S} %set of indices participating in sub-dictionary S

\safemath{\na}{n_a}			% cardinality of set of linearly independent columns of first dictionary
\safemath{\nb}{n_b}			% cardinality of set of linearly independent columns of second dictionary
\safemath{\coeffa}{p_i}	%coefficients for columns of A
\safemath{\coeffb}{q_j}	%coefficients for columns of B
\safemath{\seta}{\setP}		% set of linearly independent columns of A
\safemath{\setb}{\setQ}     % set of linearly independent columns of B
\safemath{\setw}{\setW}	%set of n largest elements of w
\safemath{\setz}{\setZ}	%set of L-n largest elements of z
\safemath{\cola}{\veca}		% generic element of the dictionary A
\safemath{\colb}{\vecb}		% generic element of the dictionary B
\safemath{\cold}{\vecd}		% generic element of the dictionary D
\safemath{\inputvec}{\vecx} 	%coefficient vector (input)
\safemath{\error}{\vece}	%error vector
\safemath{\noiseout}{\vecz} 	%noisy output vector
\safemath{\inputvecel}{x}
\safemath{\inputveca}{\vecx_a}
\safemath{\inputvecb}{\vecx_b}
\safemath{\outputvec}{\vecy}	%output of Dictionary
\safemath{\lambdamin}{\lambda_{\mathrm{min}}}
\safemath{\elltwo}{\ell_2}
\safemath{\ellone}{\ell_1}
\safemath{\ellzero}{\ell_0}
\safemath{\ellinf}{\ell_\infty}
\safemath{\ellinftilde}{\ell_{\widetilde\infty}}
\safemath{\licard}{Z(\coh,\coha,\cohb)}
\safemath{\xsol}{\hat{x}}
\safemath{\xbord}{x_b}		%Solution at the border
\safemath{\xstat}{x_s}		%Solution stationary in l0 prob
\safemath{\xstatLone}{\tilde{x}_s}
\safemath{\order}{\mathcal{O}} %order notation (big O)
\safemath{\scales}{\Theta} %scales as
\safemath{\ones}{\mathbf{1}} %all ones matrix
\safemath{\zeroes}{\mathbf{0}} %all zeroes matrix
\safemath{\thlone}{\kappa(\coh,\cohb)} %treshold l1 problem
\safemath{\constoneA}{\delta} %constant in l1 theorem to save space
\safemath{\constoneB}{\epsilon} %constant in l1 theorem to save space
\safemath{\nlarge}{L}				   %num large elements
\safemath{\sumlarge}{S_\nlarge}
\safemath{\maxlarger}{P_\nlarge}	   % maximum in Gribonval and Nielsen
\safemath{\Pzero}{\textrm{P0}}	
\safemath{\Pone}{\textrm{P1}}
\safemath{\vecfir}{\vecw}			 % \vecv element of the kernel of the dictionary, \vecv=[\vecfir \vecsec]
\safemath{\vecsec}{\vecz}
\safemath{\elvecfir}{w}              % element of vecfir
\safemath{\elvecsec}{z}				 % element of vecsec
\safemath{\nlargefir}{n}
\safemath{\normout}{\gamma}
\safemath{\auxfun}{h}
\safemath{\supp}{\textrm{supp}}%support

\safemath{\indexa}{\ell}
\safemath{\indexb}{r}
\safemath{\indexc}{i}
\safemath{\indexd}{j}

\safemath{\project}{P}%projector

\usepackage{color}

\allowdisplaybreaks % THIS ALLOWS EQUATIONS TO BREAK THE PAGE

\newcommand{\yhat}[0]{\hat{\bmy}} % y_hat for vector
\newcommand{\hhat}[0]{\hat{\bmh}} % h_hat for vector
\newcommand{\yhate}[1]{\hat{y}_{#1}} % y_hat for element
\safemath{\LAMA}{\textrm{LAMA}}
\safemath{\MRT}{\textrm{MRT}}
\safemath{\betamax}{\beta^\text{max}_\setO}
\safemath{\betamaxno}{\beta^\text{max}}
\safemath{\betamin}{\beta^\text{min}_\setO}
\safemath{\betaminno}{\beta^\text{min}}

\safemath{\Nomin}{\No^\textnormal{min}(\beta)}
\safemath{\Nominnobeta}{\No^\text{min}}
\safemath{\Nomax}{\No^\textnormal{max}(\beta)}
\safemath{\Nomaxnobeta}{\No^\textnormal{max}}
\safemath{\EX}{E_\textnormal{x}}
\safemath{\EXP}{\EX^\textnormal{p}}
\safemath{\Eo}{E_0}

\safemath{\tmax}{{t_\textnormal{max}}}
\safemath{\MAP}{\textrm{MAP}}
\safemath{\IO}{\textrm{IO}}
\safemath{\JO}{\textrm{JO}}
\safemath{\Nopost}{N_{0}^\textnormal{post}}
\safemath{\MT}{U}
\safemath{\MR}{B}
\safemath{\Tran}{\textnormal{T}}
\safemath{\Herm}{\textnormal{H}}
\safemath{\row}{\textnormal{r}}
\safemath{\col}{\textnormal{c}}

\safemath{\NT}{N_\textnormal{T}}
\safemath{\DSNR}{\delta \textnormal{SNR}}
\safemath{\betaMOR}{\beta^{\star}}
\begin{document}
	
\title{BEACHES: Beamspace Channel Estimation for Multi-Antenna mmWave Systems and Beyond}

\author{\IEEEauthorblockN{Ramina Ghods, Alexandra Gallyas-Sanhueza, Seyed Hadi Mirfarshbafan, and Christoph Studer} \\[-0.3cm]
\IEEEauthorblockA{\textit{School of Electrical and Computer Engineering, Cornell University, Ithaca, NY; email: {studer@cornell.edu}}}\thanks{The work of RG was supported by the US National Science Foundation under grants ECCS-1408006, CCF-1535897,  CCF-1652065, CNS-1717559, and ECCS-1824379. The work of AGS, SHM, and CS was supported in part by Xilinx, Inc.\ and by ComSenTer, one of six centers in JUMP, a Semiconductor Research Corporation (SRC) program sponsored by DARPA.  
}\\[-0.4cm]
}

\maketitle
\begin{abstract}
	Massive multi-antenna millimeter wave (mmWave) and terahertz wireless systems promise high-bandwidth communication to multiple user equipments in the same time-frequency resource. The high path loss of wave propagation at such frequencies  and the fine-grained nature of beamforming with massive antenna arrays necessitates accurate channel estimation to fully exploit the advantages of such systems. In this paper, we propose BEAmspace CHannel EStimation (BEACHES), a low-complexity channel estimation algorithm for multi-antenna mmWave systems and beyond. BEACHES leverages the fact that wave propagation at high frequencies is directional, which enables us to denoise the (approximately) sparse channel state information in the beamspace domain. To avoid tedious parameter selection, BEACHES includes a computationally-efficient tuning stage that provably minimizes the mean-square error of the channel estimate in the large-antenna limit. To demonstrate the efficacy of BEACHES, we provide simulation results for line-of-sight (LoS) and non-LoS mmWave channel models. 
\end{abstract}

% !TEX root = 19SPAWC_BEACHES.tex
% DO NOT REMOVE THE ABOVE COMMENT!
\section{Introduction}

Massive multiuser (MU) multiple-input multiple-output (MIMO) \cite{LarssonMMIMONextGen} and millimeter-wave (mmWave) communication~\cite{mmWillWork,mmWaveWireless} are among the key technologies of next-generation wireless  systems.
The high path loss of wave propagation at mmWave or terahertz (THz) frequencies and the fact that massive MU-MIMO enables fine-grained beamforming, requires the basestations (BSs) to acquire accurate channel state information (CSI) \cite{RappaportmmWaveModel, GaommWaveChEst}.
In addition, the trend towards low-precision data converters in all-digital massive MU-MIMO BSs to reduce power, interconnect bandwidth, and costs~\cite{ThroughputAnalysis} renders accurate channel estimation increasingly important.

At mmWave or terahertz frequencies, wave propagation is highly directional and real-world channels typically comprise only a small number of dominant propagation paths~\cite{mmWillWork,mmWaveWireless}. These unique properties enable the deployment of channel estimation algorithms that effectively suppress  noise~\cite{AlkhateebChEstHybrid2014,schniter2014channel,deng2018mmwave}.
As a consequence, compressive sensing (CS)-based methods have been proposed for mmWave channel estimation in \cite{AlkhateebLimitedFeedback, LeeOMP}.
Most of such methods use a discretization procedure of the number of propagation paths that can be resolved in the \emph{beamspace} domain~\cite{brady2013beamspace}, resulting in the well-known basis mismatch problem~\cite{tang2013compressed}. 
Methods that perform off-grid CS, such as atomic norm minimization (ANM)~\cite{BhaskarANM} or Newtonized orthogonal matching pursuit (NOMP) \cite{NewtonizedOMP}, avoid this basis mismatch problem.  These methods, however, exhibit excessively high complexity, which renders the design of corresponding hardware designs challenging.

\subsection{Contributions}
We propose a new channel estimation algorithm for massive MU-MIMO mmWave/terahertz communication systems that relies on Stein's unbiased risk estimator (SURE). Our algorithm, called BEAmspace CHannel EStimation (BEACHES), exploits sparsity of mmWave/terahertz channels in the beamspace domain and adaptively denoises the channel vectors at low complexity.
We prove that BEACHES minimizes the mean square error (MSE) of the channel estimate in the large-antenna limit.
We evaluate BEACHES for LoS and non-LoS mmWave channels and demonstrate that it performs on par with ANM and NOMP but at orders-of-magnitude lower  complexity.

\subsection{Notation}
Lowercase and uppercase boldface letters designate column vectors and matrices, respectively. 
For a vector $\bma$, the $k$th entry is~$[\bma]_k=a_k$; the real and imaginary parts are $[\bma]_{\mathcal{R}}=\bma_{\mathcal{R}}$ and $[\bma]_{\mathcal{I}}=\bma_{\mathcal{I}}$, respectively. 
The transpose and conjugate transpose of matrix $\bA$ are~$\bA^\Tran$ and~$\bA^\Herm$, respectively. 
A complex Gaussian vector~$\bma$ with mean vector~$\bmm$ and covariance matrix~$\bK$ is written as \mbox{$\bma \sim \setC\setN(\bmm,\bK)$} and its probability density function (PDF) as $f^{\setC\setN}(\bma;\bmm,\bK)$. 
A real Gaussian vector  $\bma$ with mean vector~$\bmm$ and covariance matrix~$\bK$ is written as \mbox{$\bma \sim \setN(\bmm,\bK)$} and its PDF as $f^{\setN}(\bma;\bmm,\bK)$. 
The expectation operator is $\Ex{}{\cdot}$.  
%
% !TEX root = 19SPAWC_BEACHES.tex
% DO NOT REMOVE THE ABOVE COMMENT!
\section{System Model and Beamspace Representation}
\label{sec:systemmodel}

\setlength{\textfloatsep}{8pt}
\begin{figure}[tp]
\centering
\includegraphics[width=0.85\columnwidth]{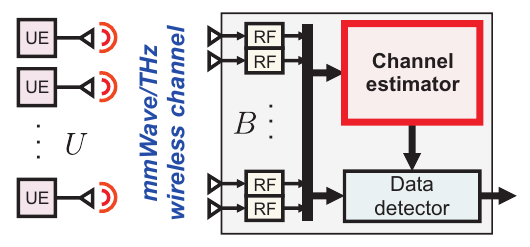}
\vspace{-0.2cm}
\caption{Massive MU-MIMO mmWave uplink system: $U$ UEs transmit pilots over a mmWave/THz wireless channel, which are used to estimate the channel vectors associated to each UE at the $B$-antenna basestation.}
\label{fig:system_overview}
\end{figure}
\subsection{System Model}
We consider an all-digital mmWave/THz massive MU-MIMO  uplink system as illustrated in \fref{fig:system_overview}. The BS is equipped with a $B$-antenna uniform linear array (ULA) and communicates with~$U$ single-antenna user equipments (UEs) in the same time-frequency resource.
For simplicity, we focus on pilot-based channel estimation for flat-fading channels, where the BS estimates the $B$-dimensional complex channel vector \mbox{$\bmh\in\complexset^B$} for each UE. 
By assuming that (i) wave propagation is predominantly directional~\cite{akdeniz2014millimeter,RappaportmmWaveModel}, and (ii) the distance between UEs and BS is sufficiently large,  the channel vectors in the antenna domain can be modeled as follows \cite{TseWireless}:
 \begin{align} \label{eq:planewavemodel}
\bmh = \sum_{\ell=0}^{L-1} \alpha_\ell \bma(\Omega_\ell), \,\, \bma(\Omega) \!=\! \big[e^{j0\Omega},e^{j1\Omega},\ldots,e^{j(B-1)\Omega} \big]^\Tran\!\!.
\end{align}
Here, $L$ refers to the total number of paths arriving at the antenna array (including a potential LoS path),  \mbox{$\alpha_\ell\in\complexset$} is the complex-valued channel gain of the $\ell$th path, and $\bma(\Omega_\ell)$ represents a complex-valued sinusoid containing the relative phases between BS antennas, where $\Omega_\ell\in[0,2\pi)$ is determined by the incident angle of the $\ell$th path to the antenna array. 
We model the estimated channel vector in the antenna domain as $\bmy=\bmh+\bme$,
where $\bme \sim \setC\setN(\bZero_{B\times1},\Eo \bI_B)$ represents channel estimation error with variance $\Eo$ per complex entry.

\subsection{Beamspace Channel Vector Denoising}
The channel vectors~$\bmh$  as modeled in~\fref{eq:planewavemodel} are a superposition of $L$ complex-valued sinusoids. Hence, it is useful to transform the vector~$\bmh$ into the discrete Fourier transform (DFT) domain, also known as the \emph{beamspace domain}, $\hhat=\bF\bmh$, where~$\bF$ is the $B\times B$ unitary DFT matrix.
In the beamspace domain, each entry of~$\hat\bmh$ is associated to a specific incident angle with respect to the BS antenna array \cite{brady2013beamspace}. 
If the number of paths~$L$ is smaller than the number of BS antennas~$B$, then the beamspace channel vector~$\hat\bmh$  will be (approximately) sparse~\cite{schniter2014channel}. 
This key property enables the use of denoising  algorithms. 
More specifically, by transforming~$\bmy$ into the beamspace domain
$\hat{\bmy}=\bF\bmy =  \hat{\bmh}+\hat{\bme}$,
where $\hat{\bme}=\bF\bme$ has the same statistics as~$\bme$, one can suppress noise while preserving the strong beamspace components. Prominent methods for beamspace denoising are ANM~\cite{BhaskarANM} and NOMP~\cite{NewtonizedOMP}, which require high complexity.
%

% !TEX root = 19SPAWC_BEACHES.tex
% DO NOT REMOVE THE ABOVE COMMENT!
\section{BEACHES: BEAmspace CHannel EStimation}
\label{sec:beaches}

\subsection{Channel Vector Denoising via Soft-Thresholding}
A widely-used sparsity-based denoising method is the least absolute shrinkage and selection operator~(LASSO)~\cite{donoho1995adapting,Tibshirani94}, which corresponds to the following optimization problem: 
\begin{align} \label{eq:LASSO}
\hhat^{\star} =\argmin_{\hhat^{\prime}\in\complexset^B} 
\textstyle \frac{1}{2} \|\yhat-\hhat^{\prime}\|_2^2+\tau \|\hhat^{\prime}\|_1.
\end{align} 
Here, $\tau\in\reals_{+}$ is a suitably-chosen denoising parameter. 
The solution to \fref{eq:LASSO} in the complex case  is the well-known \emph{soft-thresholding operator} $\eta(\yhat,\tau)$ defined as~\cite[App.~A]{maleki2013asymptotic}  
\begin{align} \label{eq:softthresholdingfunction}
[\eta(\yhat,\tau)]_b=\frac{\yhate{b}}{|\yhate{b}|}\max{\{|\yhate{b}|-\tau,0\}}, \quad b=1,\ldots,B,
\end{align}
where we define $y/|y|=0$ for $y=0$. 
For  sparsity-based denoising via soft-thresholding, the performance strongly depends on the choice of the denoising parameter~$\tau$ \cite{donoho1995adapting,mousavi2013parameterless}.
In wireless systems, it is particularly important to design robust methods to select this parameter, as many factors such as the propagation conditions, the number of arriving paths, and the signal power can vary widely over time.

\subsection{Computing the Optimal Denoising Parameter}
In what follows, we are interested in the optimal parameter~$\tau^\star$ that minimizes the estimation MSE defined as 
\begin{align} \label{eq:MSE}
\textit{MSE} = \frac{1}{B}\Ex{}{\|\hat\bmh^\star-\hhat\|_2^2}\!,
\end{align}
where $\hat\bmh^\star=\eta(\yhat,\tau^\star)$ is the associated denoised beamspace vector.
Determining the optimal parameter~$\tau^\star$ requires knowledge of the noiseless beamspace vector $\hhat$, which is unknown.  
To avoid the need for knowing the ground truth $\hhat$, we propose to use Stein's unbiased risk estimate (SURE)~\cite{donoho1995adapting} as a proxy for the MSE function.
The following result provides SURE in the complex domain and shows that it is an unbiased estimator for the MSE. The proof is given in \fref{app:MSEtoSURE}.
\begin{thm} 
\label{thm:MSEappox}
Let $\hat\bmh \in  \complexset^{B}$ be an unknown vector and \mbox{$\yhat \in \complexset^{B}$} a noisy observation vector distributed as \mbox{$\hat\bmy\sim\setC\setN(\hat\bmh,\Eo\bI_B)$}.
Let $\mu(\hat{\bmy})$ be an estimator of~$\hat\bmh$ from $\yhat$ that is weakly differentiable and operates element-wise on vectors. Then,  
\begin{align} \label{eq:complexSURE}
\textit{SURE}=&\, \frac{1}{B}\|\mu(\yhat)-\yhat\|_2^2+\Eo \nonumber \\
&\,+\frac{1}{B} \Eo \sum_{b=1}^{B} \left( \frac{\partial {[\mu_{\mathcal{R}}(\yhat)]_b}}{\partial [\yhat_{\mathcal{R}}]_b}+\frac{\partial {[\mu_{\mathcal{I}}(\yhat)]_b}}{\partial [\yhat_{\mathcal{I}}]_b}-2\right)
\end{align}
is an unbiased estimate of the MSE, i.e., $\Ex{}{\textit{SURE}} = \textit{MSE}$.
\end{thm}
The following theorem shows that SURE for the soft-thresholding operator $\eta(\yhat,\tau)$ converges to the MSE in the large-antenna limit $B\to\infty$. The  proof is given in \fref{app:shrinkageSURE}.
\begin{thm} 
	\label{thm:SURE_shrinkage}
	For the soft-thresholding function $\mu(\yhat) = \eta(\yhat,\tau)$ in \fref{eq:softthresholdingfunction}, SURE in \fref{eq:complexSURE} is given by\footnote{As discussed in \fref{app:shrinkageSURE}, the value of SURE is undefined for $\tau=\yhate{b}$, $b=1,\ldots,B$, due to the non-differentiability of the function $\eta$.}
	\begin{align} 
	\nonumber
	\textit{SURE}_\tau = &\,
	\frac{1}{B}\sum_{b:|\yhate{b}|<\tau}|\yhate{b}|^2 + \frac{1}{B}\sum_{b:|\yhate{b}|>\tau}\tau^2 + \Eo\\\label{eq:shrinkageSURE}
	&\, - \frac{\Eo}{B}\tau\sum_{b:|\yhate{b}|>\tau} \frac{1}{|\yhate{b}|} - 2\frac{\Eo}{B}\sum_{b:|\yhate{b}|<\tau}1,
	\end{align}
which, in the limit $B \to \infty$ converges to the MSE, i.e.,
\begin{align}\label{eq:SURE_convergence}
\lim\limits_{B \to \infty} \text{SURE}_\tau  = \text{MSE}.
\end{align}
\end{thm}

SURE in \fref{eq:shrinkageSURE} is \emph{independent} of the true beamspace channel vector $\hhat$. The expression \fref{eq:shrinkageSURE} only depends on the magnitudes of the observed beamspace channel vector $\yhat$, the channel estimation error variance $\Eo$, the number of BS antennas~$B$, and the denoising parameter~$\tau$. 
Thanks to \fref{eq:SURE_convergence} and the fact that~$B$ is large in massive MU-MIMO systems, we can use SURE as a surrogate to minimize the MSE and to determine the optimal denoising parameter.
While no closed-form expression for the minimum of \fref{eq:shrinkageSURE} is known, reference~\cite{mousavi2013parameterless} proposes a bisection procedure to approximate the optimal value of a similar expression for sparse recovery. We next propose an efficient algorithm that computes the optimal parameter~$\tau^\star$ using a deterministic procedure with complexity  $O(B\log(B))$.

%
% !TEX root = 19SPAWC_BEACHES.tex
% DO NOT REMOVE THE ABOVE COMMENT!

% DO NOT REMOVE, EDAS NEEDS THIS
\makeatletter
\newcommand\fs@betterruled{%
  \def\@fs@cfont{\bfseries}\let\@fs@capt\floatc@ruled
  \def\@fs@pre{\vspace*{5pt}\hrule height.8pt depth0pt \kern2pt}%
  \def\@fs@post{\kern2pt\hrule\relax}%
  \def\@fs@mid{\kern2pt\hrule\kern2pt}%
  \let\@fs@iftopcapt\iftrue}
\floatstyle{betterruled}
\restylefloat{algorithm}
\makeatother
\setlength{\textfloatsep}{5pt}
\begin{algorithm}[tp]
\caption{\strut BEACHES: BEAmspace CHannel EStimation \label{alg:BEACHES}}
\begin{algorithmic}[1]
\STATE {\bf input} $\hat\bmy = \text{FFT}(\bmy)$ and $\Eo$
\STATE  $S=0$ and $\textit{SURE}_{\text{min}}=\infty$											\label{alg:BEACHES:line2}
\STATE $\yhat^s=\text{sort}\{|\yhat|,\text{`ascend'}\}$
\STATE $V=\sum_{k=1}^{B} {(|\yhate{k}^s|)^{-1}}$, $\hat{y}^s_{0}=0$, and $\hat{y}^s_{B+1}=\infty$  	\label{alg:BEACHES:line4}
\FOR{$k=1,\ldots,B+1$}																	\label{alg:BEACHES:line5}
\STATE $\tau = \max \{ \hat{y}^s_{k-1},  \min \{ \hat{y}^s_{k}, \frac{E_0}{2(B-k+1)} V \} \}.$
\STATE $\textit{SURE}_\tau = \frac{S}{B} + \frac{(B-k+1)}{B}\tau^2 + \Eo - \frac{\Eo}{B} \tau  V -2 \frac{\Eo}{B} (k-1)$ \\[0.0cm]					\label{alg:BEACHES:line7}
\IF{$\textit{SURE}_\tau<\textit{SURE}_{\text{min}}$}													\label{alg:BEACHES:line8}
\STATE $\textit{SURE}_{\text{min}} = \textit{SURE}_\tau$
\STATE $\tau^\star = \tau$
\ENDIF																				\label{alg:BEACHES:line11}
\STATE $S = S + (\hat{y}^s_k)^2$ and $V = V - (\hat{y}^s_k)^{-1}$															\label{alg:BEACHES:line13}
\ENDFOR																				\label{alg:BEACHES:line14}
\STATE $[\hat{\bmh}^\star]_k=\frac{\hat{y}_k}{|\hat{y}_k|}\max{\{|\hat{y}_k|-\tau^\star,0\}}$, $k=1,\ldots,B$
\STATE  {\bf return} $\bmh^\star = \text{IFFT}(\hat{\bmh}^\star)$ \label{alg:BEACHES:line16}
\end{algorithmic}
\end{algorithm}

\subsection{The BEACHES Algorithm}
Reference  \cite{donoho1995adapting} outlines an efficient procedure to minimize SURE for real-valued wavelet denoising. 
We propose a similar strategy to minimize \fref{eq:shrinkageSURE} for the complex-valued case with soft-thresholding.
Instead of continuously sweeping the denoising parameter~$\tau$ in the interval $[0,\infty)$, we first sort the absolute values of the vector $\hat\bmy$ in ascending order which we call $\hat{\bmy}^s$. 
We then search for the optimal denoising parameter~$\tau$ only between each pair of consecutive elements of the sorted vector, i.e., $\tau\in\left(\hat{y}^s_{k-1}, \hat{y}^s_{k}\right)$ for $k=1,\ldots,B+1$.
In each such interval, SURE is a quadratic function in $\tau$ given by 
\begin{align}\nonumber
\textit{SURE}_{\tau} = \, &  {\sum_{b=0}^{k-1}\frac{(\hat{y}^s_{b})^2}{B}} + \frac{(B-k+1)}{B}\tau^2 + E_0 \\ &- \frac{E_0}{B}\tau {\sum_{b=k}^{B+1}(\hat{y}^s_{b})^{-1}} - 2\frac{E_0}{B}(k-1),
\label{eq:QuadraticSURE}
\end{align}
where we define $\hat{y}^s_{0}=0$ and $\hat{y}^s_{B+1}=\infty$ to account for the first $\left(0, \hat{y}^s_{1}\right)$ and last interval $\left(\hat{y}^s_{B}, \infty \right)$.
For each $k \in \{1,\ldots,B+1\}$, we compute the optimal value of $\tau$ that minimizes SURE in each interval $\tau\in\left(\hat{y}^s_{k-1}, \hat{y}^s_{k}\right)$. Since there is a discontinuity in the SURE expression when progressing from one interval to the next, the minimal value in each interval is either the minimum of the quadratic function~\fref{eq:QuadraticSURE} or one of the boundaries of the interval $\hat{y}^s_{k-1}$ and~$\hat{y}^s_{k}$.\footnote{Note that SURE is not defined for $\tau= \hat{y}^s_{k-1}$ and $\tau= \hat{y}^s_{k}$. Instead, we compute $\textit{SURE}_\tau$ for two values arbitrarily close to these boundaries, i.e., $\tau= \hat{y}^s_{k-1}+\epsilon$ and $\tau= \hat{y}^s_{k}-\epsilon$, where $\epsilon>0$ is small compared to $\tau$.}
The minimum value of~\fref{eq:QuadraticSURE} is given by 
$\tau_k^Q = \textstyle \frac{E_0}{2(B-k+1)} {\sum_{b=k}^{B+1}(\hat{y}^s_{b})^{-1}}$.
Since the function $\text{SURE}_\tau$ is convex, we can identify the optimal parameter~$\tau$ in the interval $\left(\hat{y}^s_{k-1}, \hat{y}^s_{k}\right)$ indexed by~$k$ by knowing the value of~$\tau_k^Q$ with respect to the interval boundaries $\hat{y}^s_{k-1}$ and $\hat{y}^s_{k}$.
 Put simply, the optimal denoising parameter $\tau^\star_k$ in each interval $k=1,\ldots,B+1$ is given by
\begin{align*}
\tau^\star_k = \left\{\begin{array}{ll}
\, \tau_k^Q  & \hat{y}^s_{k-1}<\tau_k^Q<\hat{y}^s_{k} \\ 
\, \hat{y}^s_{k-1} & \tau_k^Q<\hat{y}^s_{k-1} \\ 
\, \hat{y}^s_{k} & \tau_k^Q>\hat{y}^s_{k}, 
\end{array}\right.
\end{align*}
or simply 
$\tau^\star_k = \max \{ \hat{y}^s_{k-1},  \min \{ \hat{y}^s_{k}, \tau_k^Q \} \}$.
By knowing the optimal value of $\tau$ in each interval, we only need to find the minimal value of $\text{SURE}_{\tau^\star_k}$ for $k=1,\ldots,B+1$.
It is now key to realize that we do not need to recalculate SURE in~\fref{eq:QuadraticSURE} from scratch while searching through $k=1,\ldots,B+1$. Instead, we can sequentially  update the quantities $S=\sum_{b=0}^{k-1}{(\hat{y}^s_{b})^2}$ and $V=\sum_{b=k}^{B+1}{(\hat{y}^s_{b})^{-1}}$, thanks to sorting the magnitudes of the vector $\hat{\bmy}$.
The resulting procedure, called BEACHES, is summarized in \fref{alg:BEACHES}.
The computational complexity of BEACHES is only ${O}(B\log(B))$, which is caused by the FFT, sorting, and IFFT operations---the remaining computations in the for-loop (lines 5--13) are simple scalar operations.

% !TEX root = 19SPAWC_BEACHES.tex
% DO NOT REMOVE THE ABOVE COMMENT!

\setlength{\textfloatsep}{10pt}% 
\begin{figure}[tp]
\centering
\subfigure[LoS, $B=256$, $U=16$]{\includegraphics[width=0.85\columnwidth]{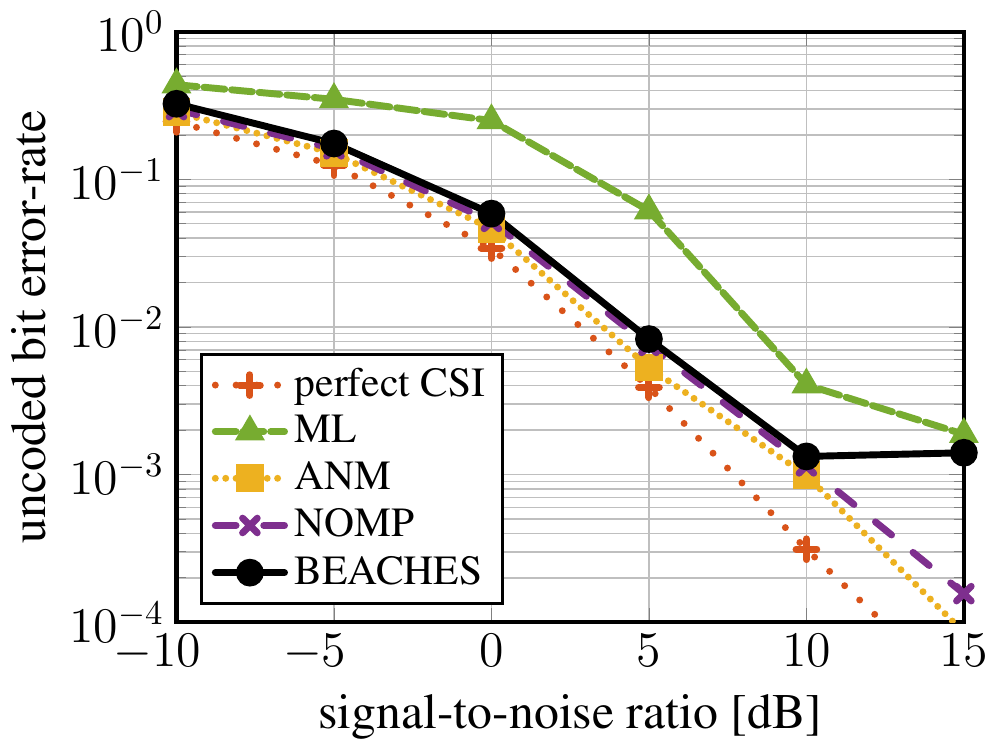}}
\hspace{1.75cm}
\subfigure[Non-LoS, $B=256$, $U=16$]{\includegraphics[width=0.85\columnwidth]{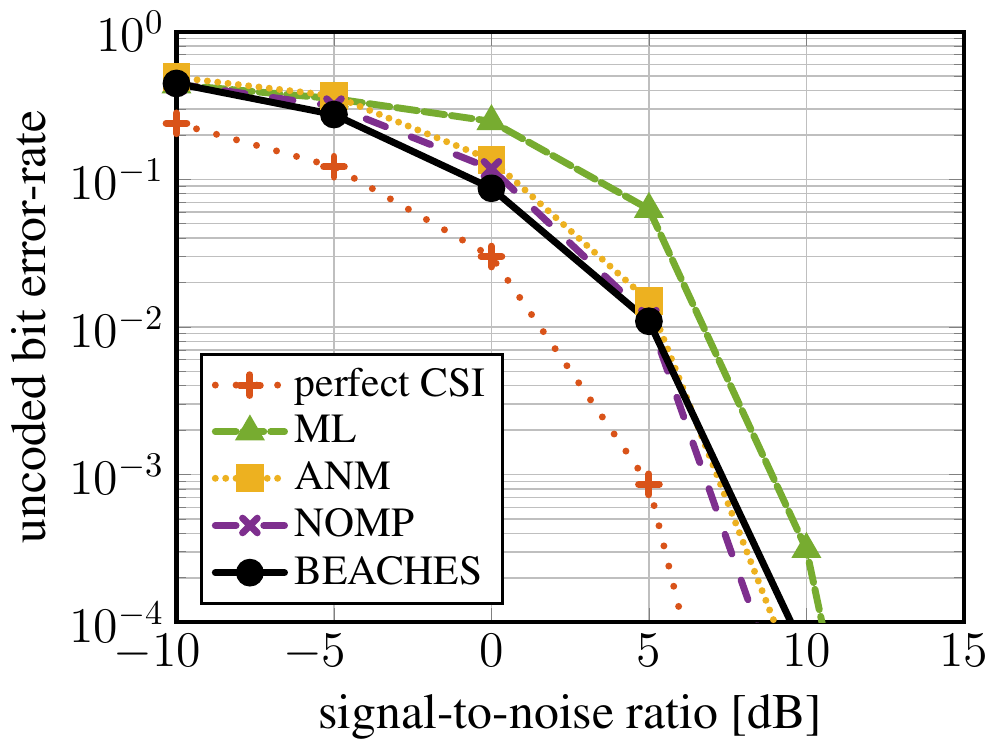}}
\vspace{-0.1cm}
\caption{Uncoded bit error-rate (BER) of various channel denoising methods for LoS and non-LoS channels. We see that BEACHES performs on par with atomic norm minimization (ANM) and Newtonized OMP, and provides $2$\,dB to $3$\,dB SNR improvements over ML channel estimation at $\textit{BER}=10^{-2}$.}
\label{fig:BER}
\end{figure}

\section{Performance and Runtime of BEACHES}
\label{sec:SimRes}

\subsection{Bit Error-Rate Performance}
To assess the performance of BEACHES, we consider an all-digital massive MU-MIMO system in which $U=16$ UEs communicate with a  $B=256$ antenna BS.
We focus on the situation in which the UEs first send orthogonal pilots, which are used to acquire maximum-likelihood channel estimates $\bmy_u$ for each UE $u=1,\ldots,U$.
The channel matrices are generated for both LoS and non-LoS conditions using the QuaDRiGa mmMAGIC UMi model~\cite{QuaDRiGa} at a carrier frequency of 60\,GHz with a ULA using $\lambda/2$ antenna spacing. 
The UEs are placed randomly within a $120^\circ$ circular sector with minimum and maximum distance of $10$ and $110$ meters from the BS antenna array, respectively. We enforce UE separation of at least~$1^\circ$ (relative to the BS antenna array) and assume optimal UE power control.
We then use different channel vector denoising methods, including 
(i) ANM-based denoising, where we use the debiased output of the code provided in~\cite{BhaskarANM},
(ii) NOMP with a (manually tuned) false alarm rate of $P_\text{fa}=0.5$ using the code provided in \cite{NewtonizedOMP},
and (iii)  ``perfect CSI,'' which is a baseline that uses the noiseless  channel vectors.
Finally, we transmit 16-QAM  symbols and perform linear minimum MSE (L-MMSE) equalization with the denoised matrix to detect the transmitted bits. 
The resulting uncoded bit error-rate (BER) is used to assess the performance of various denoising methods.

Figure~\ref{fig:BER} shows that channel vector denoising in the beamspace domain provides $2$\,dB to $3$\,dB SNR performance improvements at $\textit{BER}=10^{-2}$ compared to conventional ML channel estimation.  
The achieved performance gains are more pronounced under LoS conditions.
Quite surprisingly, we observe that BEACHES performs on par to ANM and NOMP.
This observation indicates that off-the-grid denoising methods, such as ANM and NOMP, do not provide a critical performance advantage over BEACHES (in terms of BER).

\begin{table}[tp]
	\centering
	\renewcommand{\arraystretch}{1.1}
	\begin{minipage}[c]{1\columnwidth}
		\centering
		\caption{MATLAB runtimes in milliseconds (and normalized runtimes).}
		\vspace{-0.05cm}
		\label{tbl:runtimes}
		\begin{tabular}{@{}lccc@{}}
			\toprule
			Scenario & BEACHES & NOMP & ANM \\
			\midrule
			LoS     & 1.64 (1$\times$)  & 199.9 (120$\times$)       & 47\,968 (29\,000$\times$) \\ 
			non-LoS & 1.45 (1$\times$)  & 2\,204 (1\,500$\times$) & 83\,750 (58\,000$\times$) \\						
			\bottomrule
		\end{tabular}
	\end{minipage}
\end{table}

\subsection{Runtime Comparison}
While the BER performance of BEACHES is comparable to ANM and NOMP, it exhibits significantly lower complexity. 
To support this claim, we measured their MATLAB runtimes in milliseconds on an Intel core i5-7400 CPU with 16\,GB RAM at an SNR of $5$\,dB. 
\fref{tbl:runtimes} demonstrates that the runtime of BEACHES is orders of magnitude lower than that of NOMP (up to $1\,500\times$) and ANM (up to $58\,000\times$), while the speedup is more pronounced for the non-LoS scenario.

% !TEX root = 19SPAWC_BEACHES.tex
% DO NOT REMOVE THE ABOVE COMMENT!

\section{Conclusions}
\label{sec:conclusions}
We have proposed a new channel denoising  algorithm  
for massive MU-MIMO mmWave and terahertz communication systems called BEAmspace CHannel EStimation (BEACHES).
BEACHES exploits sparsity of mmWave/terahertz channels in the beamspace domain  to perform adaptive soft-thresholding via Stein's unbiased risk estimate (SURE).
We have shown that BEACHES minimizes the mean square error in the large-antenna limit and performs on par with sophisticated channel estimation algorithms for realistic LoS and non-LoS channel models but at orders-of-magnitude lower complexity. 
There are many avenues for future work. An extension of BEACHES to systems with low-precision quantizers and single-carrier transmission is an open research problem. 
\appendices 
% !TEX root = 19SPAWC_BEACHES.tex
% DO NOT REMOVE THE ABOVE COMMENT!
\section{Proof of \fref{thm:MSEappox}}
\label{app:MSEtoSURE}
The MSE  for $\hhat^\star=\mu(\yhat)$ is defined as
\begin{align*}
\textstyle \textit{MSE} = \Ex{}{\frac{1}{B}\|\hhat^\star-\hhat\|_2^2}=\Ex{}{\frac{1}{B}\|\mu(\yhat)-\hhat\|_2^2}\!,
\end{align*}
where we decompose the complex-valued vector $\yhat$ into the real part $\yhat_{\mathcal{R}} \sim \setN(\bmh_{\mathcal{R}},\textstyle\frac{\Eo}{2} \bI_B)$ and imaginary part $\yhat_{\mathcal{I}} \sim \setN(\bmh_{\mathcal{I}},\textstyle\frac{\Eo}{2} \bI_B)$. Note that expectation is with respect to the noisy observation~$\yhat$.
Define $g(\yhat)=\mu(\yhat)-\yhat$. Hence,
\begin{align}\label{eq:MSE_complex}
\textit{MSE}=&\, \textstyle \frac{1}{B} \Ex{}{\|g(\yhat)+\yhat-\hhat\|_2^2} \nonumber \\  
= &\,\textstyle  \Ex{}{\frac{1}{B}\|g(\yhat)\|_2^2} + \Ex{}{\frac{1}{B}\|\yhat-\hhat\|_2^2} \nonumber \\ 
&\, \textstyle +\Ex{}{\frac{2}{B}\left[g(\yhat)^\Herm(\yhat-\hhat)\right]_{\mathcal{R}}}\!.
\end{align}
The last term can be simplified as 
\begin{align*}
\textstyle\frac{2}{B}\Ex{}{\left[g(\yhat)^\Herm(\yhat\!-\!\hhat)\right]_{\mathcal{R}}} \!= &\,
\textstyle\frac{2}{B}\Ex{}{{g_{\mathcal{R}}(\yhat)^\Tran(\yhat_{\mathcal{R}}\!-\!\hhat_{\mathcal{R}})}} \nonumber \\
&\textstyle +\frac{2}{B}\Ex{}{{g_{\mathcal{I}}(\yhat)^\Tran(\yhat_{\mathcal{I}}\!-\!\hhat_{\mathcal{I}})}}\!.
\end{align*}
We can now expand $\frac{2}{B}\Ex{}{{g_{\mathcal{R}}(\yhat)^\Tran(\yhat_{\mathcal{R}}-\hhat_{\mathcal{R}})}}$ which yields
\begin{align}
\textstyle \frac{2}{B}&\Ex{}{{g_{\mathcal{R}}(\yhat)^\Tran(\yhat_{\mathcal{R}}-\hhat_{\mathcal{R}})}} \\ \nonumber
\textstyle \stackrel{\text{(a)}}{=} &\, \textstyle \frac{2}{B} \int_{\yhat_{\mathcal{I}}} f^{\setN}\left(\yhat_{\mathcal{I}};\hhat_\mathcal{I},\frac{\Eo}{2} \bI_B\right) \sum_{b=1}^{B} \int_{\yhat_{\mathcal{R}}} \frac{1}{\left(2 \pi \frac{\Eo}{2}\right)^{B/2}} \times\nonumber  \\
&\, \textstyle \exp{\!\left(-\frac{\|\yhat_{\mathcal{R}}-\hhat_{\mathcal{R}}\|^2}{2\frac{\Eo}{2}}\right)}  \frac{\Eo}{2} \frac{\partial {[g_{\mathcal{R}}(\yhat)]_b}}{\partial [\yhat_{\mathcal{R}}]_b} d \yhat_{\mathcal{R}} d \yhat_{\mathcal{I}}  \\
=&\, \textstyle \frac{\Eo}{B} \Ex{}{\sum_{b=1}^{B} \left(\frac{\partial {[\mu_{\mathcal{R}}(\yhat)]_b}}{\partial [\yhat_{\mathcal{R}}]_b}-1\right)}\!, 
\label{eq:ReRe}
\end{align}
where $\text{(a)}$ follows from integration by parts. Similarly, we have
\begin{align}\label{eq:ImIm}
\textstyle\frac{2}{B}\!\Ex{}{{g_{\mathcal{I}}(\yhat)^\Tran(\yhat_{\mathcal{I}}\!-\!\hhat_{\mathcal{I}})}} 
\!=\! \frac{\Eo}{B} \!\Ex{}{ \sum_{b=1}^{B} \!\left(\frac{\partial {[g_{\mathcal{I}}(\yhat)]_b}}{\partial [\yhat_{\mathcal{I}}]_b}\!-\!1\right)}\!.
\end{align}
Recall that $g(\yhat)=\mu(\yhat)-\yhat$ and replace \fref{eq:ReRe} and \fref{eq:ImIm} in the original MSE expression in~\fref{eq:MSE_complex}. This leads to
\begin{align*}
\textstyle \textit{MSE}= &\, \textstyle \Ex{}{ \frac{1}{B}\|\mu(\yhat)-\yhat\|_2^2}+\Ex{}{ \frac{1}{B}\|\yhat-\hhat\|_2^2} \nonumber \\ 
& \textstyle +\frac{\Eo}{B} \Ex{}{ \sum_{b=1}^{B} \left( \frac{\partial {[\mu_{\mathcal{R}}(\yhat)]_b}}{\partial [\yhat_{\mathcal{R}}]_b}+\frac{\partial {[\mu_{\mathcal{I}}(\yhat)]_b}}{\partial [\yhat_{\mathcal{I}}]_b}-2\right)}\!.
\end{align*}
The second term in the MSE expression above equals $\Eo$. For the first and third term we remove their expectations to arrive at the following SURE expression:
\begin{align*}
\textstyle\textit{SURE}=&\, \textstyle \frac{1}{B} \|\mu(\yhat)-\yhat\|_2^2+\Eo \nonumber \\ 
&\, \textstyle +\frac{\Eo}{B} \sum_{b=1}^{B} \left( \frac{\partial {[\mu_{\mathcal{R}}(\yhat)]_b}}{\partial [\yhat_{\mathcal{R}}]_b}+\frac{\partial {[\mu_{\mathcal{I}}(\yhat)]_b}}{\partial [\yhat_{\mathcal{I}}]_b}-2\right)\!,
\end{align*}
which establishes the fact that $\Ex{}{\textit{SURE}} = \textit{MSE}$.

% !TEX root = 19SPAWC_BEACHES.tex
% DO NOT REMOVE THE ABOVE COMMENT!
\section{Proof of \fref{thm:SURE_shrinkage}}
\label{app:shrinkageSURE}
SURE in \fref{eq:complexSURE} for $\mu(\hat{\bmy})=\eta(\yhat,\tau)$ is derived as follows. The only unknowns in the expression of SURE are its derivative of real and imaginary parts.
For $|\hat{y}_b|<\tau$, we have
\begin{align*}
\textstyle \frac{\partial {[\eta_{\mathcal{R}}(\hat{\bmy},\tau)]_b}}{\partial [\hat{\bmy}_{\mathcal{R}}]_b}=\frac{\partial {[\eta_{\mathcal{I}}(\hat{\bmy},\tau)]_b}}{\partial [\hat{\bmy}_{\mathcal{I}}]_b}=0.
\end{align*}
For $|\hat{y}_b|>\tau$, we have
\begin{align*}
\textstyle \frac{\partial {[\eta_{\mathcal{R}}(\hat{\bmy},\tau)]_b}}{\partial [\hat{\bmy}_{\mathcal{R}}]_b}  =  &\, \textstyle \frac{\partial }{\partial [\hat{\bmy}_{\mathcal{R}}]_b} \Big([\hat{\bmy}_{\mathcal{R}}]_b-\frac{[\hat{\bmy}_{\mathcal{R}}]_b \tau}{\sqrt{[\hat{\bmy}_{\mathcal{R}}]_b^2+[\hat{\bmy}_{\mathcal{I}}]_b^2}}\Big) \nonumber \\ 
= &\, \textstyle 1-\tau \frac{[\hat{\bmy}_{\mathcal{I}}]_b^2}{{([\hat{\bmy}_{\mathcal{R}}]_b^2+[\hat{\bmy}_{\mathcal{I}}]_b^2)}^{3/2}}
\end{align*}
and
\begin{align*}
\textstyle \frac{\partial {[\eta_{\mathcal{I}}(\hat{\bmy},\tau)]_b}}{\partial [\hat{\bmy}_{\mathcal{I}}]_b} = &\, \textstyle 1-\tau \frac{[\hat{\bmy}_{\mathcal{R}}]_b^2}{{([\hat{\bmy}_{\mathcal{R}}]_b^2+[\hat{\bmy}_{\mathcal{I}}]_b^2)}^{3/2}}.
\end{align*}
Note that at $|\hat{y}_b|=\tau$, there is a discontinuity and thus the derivative and consequently SURE are not defined for this value.
The complex-valued SURE expression reduces to
\begin{align*}
\textit{SURE}_\tau = &\, \textstyle \frac{1}{B}\sum_{b=1}^{B}{\min\{|\hat{y}_b|,\tau\}}^2 +\Eo \nonumber \\ 
&\, \textstyle +\frac{\Eo}{B} \sum_{b:|\hat{y}_b|>\tau}\Big(2-\tau \frac{1}{\sqrt{[\hat{\bmy}_{\mathcal{R}}]_b^2+[\hat{\bmy}_{\mathcal{I}}]_b^2}}-2\Big)  \nonumber \\
&\, \textstyle +\frac{\Eo}{B} \sum_{b:|\hat{y}_b|<\tau}\left(0-2\right)\!.
\end{align*}

We now prove the convergence of SURE in \fref{eq:SURE_convergence}. 
In \cite[Lemma 4.3.]{mousavi2015consistent}, the authors prove convergence of SURE to MSE in the real domain for the soft-thresholding function. We follow the same procedure for the complex domain. 
Using \cite[Thm. \uppercase\expandafter{\romannumeral 3}.15 \& \uppercase\expandafter{\romannumeral 3}.16]{maleki2013asymptotic}, we have that for any pseudo-Lipschitz function $\gamma: \complexset \to \reals$ the following equality holds:
\begin{align}\label{eq:lipschitz_stuff}
& \textstyle \lim\limits_{B \to \infty} \frac{1}{B} \sum_{b=1}^{B}\gamma(\eta(\hat{y}_b,\tau),\hat{h}_b) \notag \\ 
& \textstyle \qquad \qquad =  \Ex{}{\gamma(\eta(H+\sqrt{\Eo}Z,\tau),H)}\!.
\end{align}
Here, $Z \sim \setC \setN(0,1)$ and $H$ is a random variable with the sparse distribution of the channel vector in the beamspace domain $\hat{h}_b$. 
Using \fref{eq:lipschitz_stuff}, we have the following result
\begin{align*}
\textstyle \lim\limits_{B \to \infty} \frac{1}{B} \sum_{b=1}^{B} |\eta(\hat{y}_b,\tau)-\hat{y}_b|^2 = \Ex{\hat{y}_{\tilde{b}}}{|\eta(\hat{y}_{\tilde{b}},\tau)-\hat{y}_{\tilde{b}}|^2}\!,
\end{align*}
where, $\hat{y}_{\tilde{b}}$ is any element of the random vector $\yhat$. The expression above can be rewritten as 
\begin{align}\label{eq:term1}
\textstyle \lim\limits_{B \to \infty} \frac{1}{B} \|\eta(\hat{y}_b,\tau)-\hat{y}_b\|_2^2 = \Ex{\yhat}{\frac{1}{B} \|\eta(\yhat,\tau)-\yhat\|_2^2}\!.
\end{align}
Now, since $\frac{\partial {[\eta_{\mathcal{R}}(\hat{\bmy},\tau)]_b}}{\partial [\hat{\bmy}_{\mathcal{R}}]_b}+\frac{\partial {[\eta_{\mathcal{I}}(\hat{\bmy},\tau)]_b}}{\partial [\hat{\bmy}_{\mathcal{I}}]_b}$ is bounded, it is pseudo-Lipschitz and, hence, we can use \fref{eq:lipschitz_stuff} to obtain the following convergence result:
\begin{align}\label{eq:term3}\nonumber
&\textstyle \lim\limits_{B \to \infty} \frac{1}{B} \sum_{b=1}^{B} \left( \frac{\partial {[\mu_{\mathcal{R}}(\yhat)]_b}}{\partial [\yhat_{\mathcal{R}}]_b}+\frac{\partial {[\mu_{\mathcal{I}}(\yhat)]_b}}{\partial [\yhat_{\mathcal{I}}]_b}-2\right) \\
& \textstyle\quad = \frac{1}{B} \Ex{}{ \sum_{b=1}^{B} \left( \frac{\partial {[\mu_{\mathcal{R}}(\yhat)]_b}}{\partial [\yhat_{\mathcal{R}}]_b}+\frac{\partial {[\mu_{\mathcal{I}}(\yhat)]_b}}{\partial [\yhat_{\mathcal{I}}]_b}-2\right)}\!.
\end{align}
By summing \fref{eq:term1} and \fref{eq:term3} and $\Ex{}{ \frac{1}{B}\|\yhat-\hhat\|_2^2} = \Eo$, we have established that $\lim_{B \to \infty} \text{SURE}_\tau = \text{MSE}$.

\balance
\bibliographystyle{IEEEtran}
\bibliography{VIPabbrv,publishers,confs-jrnls,REFs,VIP,ref2}
\balance

\end{document}